\documentclass[crop]{CSL}

\usepackage{braket}

\usepackage{graphicx}
\newcommand{\beq}{\begin{equation}}
\newcommand{\beqa}{\begin{eqnarray}}
		  \newcommand{\eeq}{\end{equation}}
\newcommand{\eeqa}{\end{eqnarray}}
\newcommand{\lsim}{\lesssim}
\newcommand{\gsim}{\gtrsim}

\newcommand{\lmk}{\left(}
\newcommand{\rmk}{\right)}

\newcommand{\lla}{\left\langle}

\newcommand{\rra}{\right\rangle}

\newcommand{\vp}{\varpi}

\begin{document}

\supertitle{Research Paper}

\title[]{{Astrometric study of Gaia DR2 stars for interstellar communication}}

\author[N. Seto and K. Kashiyama]{Naoki Seto$^{1}$, Kazumi Kashiyama$^{2,3}$}

\address{\add{1}{Department of Physics, Kyoto University, Kyoto 606-8502, Japan,} 
\add{2}{Research Center for the Early Universe, Graduate School of Science, University of Tokyo, Bunkyo-ku, Tokyo 113-0033, Japan,} 
 \add{3}{Department of Physics, Graduate School of Science, University of Tokyo, Bunkyo-ku, Tokyo 113-0033, Japan}}


\begin{abstract}
We discuss the prospects of high precision pointing of our transmitter to habitable planets around  Galactic main sequence  stars.
For an efficient signal delivery, the future sky positions of the host stars should be appropriately  extrapolated with accuracy better than the beam opening  angle $\Theta$ of the transmitter.  
Using the latest data release (DR2) of Gaia, we estimate the accuracy of the extrapolations individually for $4.7\times 10^7$ FGK stars, and find that the total number of targets could be $\sim 10^7$ for the accuracy goal better than 1\rq\rq{}. Considering the   pairwise nature of communication, 
our study would be instructive also for SETI (Search for Extraterrestrial Intelligence), not only for sending signals outward.
\end{abstract}


\selfcitation{}{}

\received{xx xxxx xxxx}

\revised{xx xxxx xxxx}

\accepted{xx xxxx xxxx}

\maketitle

\Fpagebreak

\vfill\pagebreak

\section{Introduction}

Even $\sim60$ years have passed since the pioneering work by Drake, we have not succeeded to detect a convincing signature of extraterrestrial intelligence (ETI) (e.g., Drake 1961; Horowitz, \& Sagan 1993; Tarter 2001; Siemion et al. 2013).   On one hand, this might be simply reflecting the possibility that the number of  Galactic civilizations is small, or even  zero on  our past light-cone.  On the other hand, our observational facilities and available computational resources might not be sufficient to deal with existing weak signals in a huge parameter space (Tarter et al. 2010; Wright et al. 2018).  In any case,  SETI programs are actively ongoing, including recently launched Breakthrough Listen in which $\sim 10^6$ Galactic stars and $\sim 10^2$ nearby galaxies will be analyzed (Gajjar et al. 2019).

In parallel with the searching efforts, artificial signals have been intentionally transmitted from the Earth to extraterrestrial systems (e.g., Zaitsev 2016, see also Baum et al. 2011; Vakoch 2016). For example, Polaris has been repeatedly selected as a target. In 2008, a 70~m-dish antenna of NASA\rq{}s Deep Sky Network was used for a radio transmission in the X-band (wavelength $\lambda\sim 4$~cm).  More recently, in 2016,  an ESA\rq{}s antenna (dish size $L=35$~m) was directed to Polaris for sending messages encoded in $\lambda\sim 4$~cm radio  waves.  Here we should comment that the 1-10~GHz band (wavelength 3-30~cm) is regarded  as an ideal window for interstellar communication, given the background noises (Cocconi \& Morrison 1959). The half opening angle of the transmitted beam is given by $\Theta\sim \lambda/(2L)$, and we have $O$(100") for the two concrete cases mentioned above. 

Considering the potential limitations of observational facilities and computational resources inversely at  ETI side, it  would be more advantageous to increase the energy flux of our outgoing signals.  Here one of the solid  options is to reduce the beam opening angle $\Theta$ (e.g., Benford et al. 2010, see also  Hippke 2019). For example, using a phased array with an effective diameter comparable to the core station of SKA2, we can realize $\Theta \sim 0.8" (\lambda/{\rm 4~cm})(L/{\rm 5~km})^{-1}$.   
Note that, for a given beam opening angle $\Theta$,  the size of the transmitter $L$ could be  reduced by using a shorter wavelength $\lambda$. 
Clark \& Cahoy (2018) studied intentional  signal transmissions in the optical/IR bands for which typical seeing level on the surface of the earth is $O(1")$. 
Even though the Sun becomes a much stronger background than in the radio band, they discussed that a facility similar to the Airborne Laser ($L=1.5$~m, $\lambda=1315$~nm, $\Theta\sim 0.1"$) could be workable, depending on the size of the receiver\rq{}s 
telescope. 
Meanwhile, the lightsail propulsion has been studied as an attractive technology for future interplanetary and interstellar missions. Under certain restrictions, Guillochon \& Loeb (2015) showed an optimal combination $L=1.5$~km and $\lambda=0.4$~cm for interplanetary transportations, corresponding to $\Theta\sim 0.3"$.  With a potential light beamer ($L\sim1~$km, $\lambda\sim 1~\mu$m) for the Breakthrough Starshot, the diffraction would be $O(0.001")$. \footnote{https://breakthroughinitiatives.org/forum/28?page=4}

However, a target star is moving on the sky with proper motion $\mu$, and there is an offset angle between its observed position and the appropriate transmission direction. Given the round-trip time $2d/c$ ($d$: the target distance) of photon, the offset is estimated to be  $2d/c\times \mu=2(v_{\rm t}/c)=40"(v_{\rm t}/{\rm 30~km\, s^{-1}}$) with the transverse velocity $v_{\rm t}$ (Arnold 2013; Zaitsev 2016).
Therefore, if we use a beam  $\Theta \lsim 40$\rq\rq{} and want to shoot a star moving at the typical transverse velocity $v_{\rm t} \sim {\rm 30~km\, s^{-1}}$ (De Simone et al. 2004), we generally need to carefully extrapolate the future position of the star,  by measuring its related parameters.  For example, we   
require the precision  $\Delta v_{\rm t}\sim 0.75{\rm ~km\, s^{-1}} (\Theta/1")$ for the transverse velocity.

An astrometric mission is an ideal instrument for this measurement.  It provides us with five astrometric  parameters: the sky position $(\alpha,\delta)$, parallax $\vp$ and proper motion $(\mu_\alpha,\mu_\delta)$.  All of them are indispensable  for our extrapolation.

In 2016, the astrometric mission Gaia released its first data (DR1). 
Since then, Gaia has brought significant impacts on various fields of astronomy.  Its unprecedented precision is expected to also change the shooting problem drastically.

In this paper, using Gaia\rq{}s latest data release (DR2, Brown et al. 2018), we estimate the number of main-sequence FGK stars suitable for the high precision shooting. These stars would have their habitable zones at $\sim 1$~AU. If we observe the systems at the distances of  $d$, the angular separation between the stars and their habitable planets are smaller than $0.1"(d/10~{\rm pc})^{-1}$. Therefore, we can quite certainly hit the habitable planets once its host star is within our beam of $\Theta \gsim 0.1"(d/10~{\rm pc})^{-1}$.
Note that $\sim 20$\% of the Galactic Sun-like  stars could have Earth-size planets in their habitable zones (Petigura et al. 2013), but Gaia is unlikely to detect these small planets by astrometric drift (Perryman et al. 2014).

We expect that our study would be useful also for SETI, not only for sending signals outward. This is because, it would be advantageous for receivers  to inversely assess the potential criteria and strategies  of senders at selecting their targets  
(see e.g., Schelling 1960; Wright 2018; Seto 2019).

\section{Astrometric observation and extrapolation}

\begin{figure*}[]
\vspace{8mm}
 \begin{center}
 \centerline{ \includegraphics[width=2.5in,scale=1,angle=270]{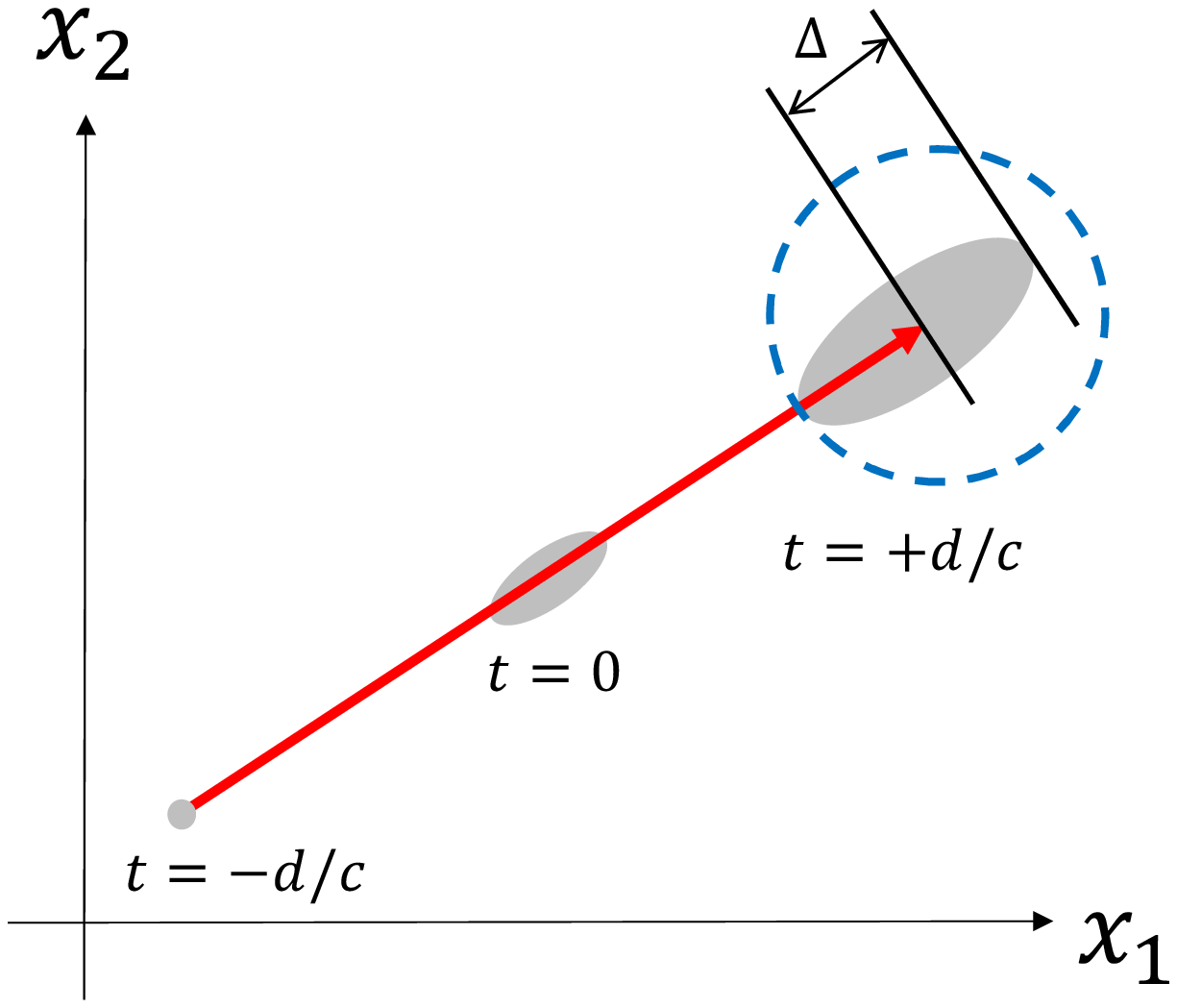}}
   \caption{ Prediction of the sky position of a target star in an orthogonal
angular coordinate in the celestial sphere. We obtain astrometric information
(sky position, parallax and  proper motion) of the target at $t = -d/c$ ($d$:
the parallax distance to the target). We then extrapolate the sky position of
the target at the hitting epoch $t = d/c$. The directional errors are shown
with gray regions. Because of the inaccuracies of the distance and the proper
motion, the error size at $t = d/c$ would be much larger than the original size
at $t= -d/c$. We define  $\Delta$ as the angular size of the long axis of the error ellipsoid. To hit the target star at a single shot, we  require that $\Delta$ is smaller than the beam width $\Theta$
of the transmitter indicated by the radius of the dashed blue  circle.  }
 \end{center}
\end{figure*}

Here we briefly explain the basic astrometric parameters relevant to our study. On the celestial sphere around a target star, we locally introduce an orthogonal angular coordinate $(x_1,x_2)$ (see Fig. 1).  The parallax $\vp$ is the annual positional modulation on the plane and is determined by  the distance $d$ to the star as
\beq
\vp=1.0  \lmk\frac{d}{\rm 1~kpc}  \rmk^{-1}  {\rm mas}.
\eeq 
 The proper motion $(\mu_1,\mu_2)$ corresponds to the long-term signature of the time derivatives $\mu_{\rm i}=dx_{\rm i}/dt$ (i$=1,2$) and is related to the 
transverse  velocity components  as  
\beq
v_{\rm t,i}=\mu_{\rm i} \times d={4.8} \lmk  \frac{\mu_{\rm i}}{\rm  1mas~yr^{-1}} \rmk \lmk  \frac{d}{\rm 1~kpc} \rmk{\rm~ km~sec^{-1}}.\label{v}
\eeq
We put $v_{\rm t}\equiv (v_1^2+v_2^2)^{1/2}$ for the magnitude of the transverse velocity.

Next we discuss the sky position of the target star for our transmission.  
We introduce a simple Galactic-scale time coordinate $t$.  Also for simplicity, we assume that the astrometric measurement and the shooting are done at the same time $t=0$ on the Earth.
But it is straightforward to incorporate the time interval (realistically $\ll d/c$) between the two operations. 

By an astrometric observation, we can basically obtain information of the target at $t=-d/c$ (see Fig. 1). In contrast, our signal strikes  the target around $t=d/c$.  Therefore,  the following extrapolation is required for the shooting 
\beqa
x_{\rm ie}&=&x_{\rm im}+ \frac{{2d_{\rm m}} \mu_{\rm im} \label{ex}}c\\
 &=& x_{\rm im}+\frac{f \mu_{\rm im}}{\vp_{\rm m}} \label{exu}
\eeqa
(see Fig. 1).   Here we use the subscript \lq\lq{}m\rq\rq{} for the astrometrically remeasured values (e.g., $x_{\rm im}\equiv x_{\rm i}(t=-d/c)$) and \rq\rq{}e\rq\rq{} for the extrapolated values for the target (e.g., $x_{\rm ie}\equiv x_{\rm i}(t=d/c)$). 
If we use the unit [mas\, yr${^{-1}}$] for $\mu_{\rm im}$ and [mas] for   $(x_{\rm ie},x_{\rm im})$, we have the numerical value $f={\rm (2~kpc/1light~year)}=6520$.

As shown in Eq. (\ref{v}),   the transverse velocity  is given by the product $d_{\rm m} \mu_{\rm im}=v_{\rm t, im}$ and is the primary quantity for adjusting the shooting direction. More specifically, as mentioned earlier and shown in Eq. (\ref{ex}), the offset angle for the shooting is given by $2(v_{\rm t,im}/c)$.  But, in standard astrometric observations, we separately estimate $d_{\rm m}\propto 1/\vp_{\rm m} $ and $\mu_{\rm im}$.

From Eq. (\ref{exu}),
the error for the extrapolated position $x_{\rm ie}$  is given by 
\beq
\delta x_{\rm ie}=\delta x_{\rm im}+f\frac{ \delta \mu_{\rm im}}{\vp_{\rm m}} -f\frac{  \mu_{\rm im} \delta \vp_{\rm m}  }{\vp_{\rm m}^2} \label{e3}.
\eeq
The first term represents the directional error of the target at $t=-d/c$. The second and third terms are those associated with the extrapolation and caused by  the errors for the proper motion and the parallax respectively.

The total number of  Gaia DR2 sources is 1,692,919,135 (Brown et al. 2018). Among them, the five astrometric parameters $(\vp,\alpha,\delta, \mu_{\alpha*},\mu_\delta)$ are provided for 1,331,090,727 sources, accompanied by the estimation of the associated $5\times 5$ error matrix.   Using these data, we can evaluate the $2\times 2$ matrix $A$ for the directional error $\delta x_{\rm ie}$ 
\beq
A\equiv \left(
    \begin{array}{cc}
      \lla  \delta x_{\rm 1e} \delta x_{\rm 1e}\rra  &  \lla  \delta x_{\rm 1e} \delta x_{\rm 2e}\rra  \\
     \lla  \delta x_{\rm 1e} \delta x_{\rm 2e}\rra  & \lla  \delta x_{\rm 2e} \delta x_{\rm 2e}\rra 
    \end{array}
  \right). \label{mat}
\eeq
This matrix determines the error ellipse  of the  extrapolated position  $(x_{\rm 1e}, x_{\rm 2e})$ in the sky, and we define $\Delta$ as the angular size of  its long axis (given in terms of  the larger eigenvalue of $A$).  If we use an transmitter with a beam opening angle $\Theta$,  we should have $\Delta<\Theta$ for hitting the target at a single shot (see Fig. 1).

Our discussions on the uncertainty $\Delta$  have been somewhat abstract. Here we make an order-of-magnitude estimation for $\Delta$. 
For an astrometric observation like Gaia, we have  approximate relations for the estimation errors of the related  parameters (e.g. $\sigma_\vp\equiv \lla \delta \vp \delta \vp\rra^{1/2}$) as
\beq
\sigma_\vp\sim \sigma_{\rm xi}\sim \sigma_{\rm \mu i}
\eeq
in the units mentioned after Eq. (\ref{exu}) (Brown et al. 2018). Therefore, in Eq. (\ref{e3}), the first term is $\sim  10^4(d/1{\rm ~kpc})$ times smaller than the second one, and is negligible for our targets at $d=O({\rm 1~kpc })$ discussed in the next section.  The ratio between the second and third terms in Eq. (\ref{e3}) is given by $(v_{\rm t}/4.8{\rm ~ km \, s^{-1}})$ with the transverse velocity $v_{\rm t}$.  For its typical value $v_{\rm t}\sim30 {\rm ~km \, s^{-1}}$, Eq. (\ref{e3}) is dominated by the third term due to the parallax  (distance) error, and we have 
\beqa
\Delta &\sim& 30"\lmk \frac{v_{\rm t}}{30 {\rm ~km \, s^{-1}}}\rmk    \lmk \frac{\sigma_\vp}{\vp}\rmk \\
&\sim& 3" \lmk \frac{v_{\rm t}}{30 {\rm ~km \, s^{-1}}}\rmk  \lmk \frac{d}{\rm ~1~kpc}\rmk   \lmk \frac{\sigma_\vp}{0.1 {\rm~ mas}}\rmk .
\eeqa
In  Eq. (9), we used the characteristic value  $\sigma_\vp\sim 0.1 {\rm ~mas}$ of Gaia DR2   for a star at G-magnitude $G=17$ (Brown et al. 2018).
In this manner, we can roughly estimate the angular uncertainty  $\Delta=O(1")$ for Gaia DR2 sources at distances $d=O(1)$~kpc.

So far, we simply fixed the target distance at the observed value $d=d_{\rm m}$  without considering its time variation. We can, in principle, measure the line-of-sight velocity $v_l$ of the target.    But, the line-of-sight velocity $v_l$ introduces an effective  change of  the transverse velocity only  by $O(v_l/c)$ and is totally negligible, compared with the required accuracy level $ \sim0.75 {\rm~km\, s^{-1}}(\Theta/1")$.  
Meanwhile the acceleration of the solar system  is estimated to be $O({\rm 10~mm\,sec^{-1}yr^{-1}})$ and is dominated by the Galactic centrifugal acceleration  (see e.g., Titov, \& Lambert 2013). If we  regard this as the typical secular value for Galactic field stars, the effective velocity shift becomes $0.07{\rm ~km\,sec^{-1}}(d/{\rm 1~kpc})$ and is not important for the accuracy goal $O(1")$. In addition, the Galactic potential could be modeled relatively well. Below certain accuracy level, it would be required to deal with  additional time-depending   fluctuations of the photon rays such as the relativistic corrections   and non-vacuum effects (e.g., scintillation). We leave related studies as our future works.

\section{Target stars in Gaia DR2}

In  this section, we examine the actual data set provided in Gaia DR2 and estimate the numbers 
 of stars suitable for our high-precision shooting.

\begin{table*}[t!]
\begin{center}
\caption{Numbers of our filtered  sample and shooting targets}
\begin{tabular}{c|c|c|c|c}
& filtered sample & targets $\Delta<5$\rq\rq{}& targets $\Delta<1$\rq\rq{}  &targets $\Delta <0.1$\rq\rq{}\\
\hline
F&  2960592& 2320725& 888155  & 8065   \\
G& 14224687  & 9183380& 2603038  & 21441   \\
 K& 30251412 & 20315244& 4819918  & 80339 \\
\hline
 total& 47436691 &31819349 &  8311111& 109945
\end{tabular}
\end{center}
\end{table*}

\subsection{FGK-type stars}

Gaia DR2 contains 76,956,778 sources  whose effective temperature $T_{\rm eff}$ and radii $R$ are presented, in addition to  the five  astrometric parameters and their $5\times 5$ noise covariance matrix.\footnote{According to Gaia DR2 site, the uncertainties are underestimated by 7-10\% for faint sources  with $G>16$ outside the Galactic plane, and by up to $\sim$ 30 per cent for bright stars with $G<12$.}  From these sources, we further  selected FGK stars potentially hosting habitable planets, by applying the following three filters;  

(i) effective temperature in the  three ranges below, F-type: $T_{\rm eff}\in $(6000~K, 7500~K], G-type:  $T_{\rm eff}\in $(5200~K, 6000~K] and  K-type :$T_{\rm eff}\in $[3700~K, 5200~K],

(ii) stellar radius: $R\le 2.0~R_\odot$  for F-type stars and $R\le 1.5~R_\odot$ for GK-type stars, 

(iii) \lq\lq{}Priam
flag\rq\rq{} value either of the following ones:  0100001, 0100002, 0110001, 0110002, 0120001 and 0120002.\par
\noindent
The filter (ii) is for removing evolved stars, and (iii) is for excluding low quality data  (Andrae et al. 2018).

After the selection, we obtained $4.7\times 10^7$ stars, as shown in the first column in Table 1. In the following, we call these stars \lq\lq{}filtered sample\rq\rq{}. 
 They are anisotropically distributed in the sky with the averaged density $\rm \sim10^{-4} [arcsec^{-2}]$.   
In Fig. 2 (cyan curve), we present the G-magnitude distribution of our filtered sample. We have a sharp cut-off at $G=17$ that is mainly determined by  the availability of the effective temperature $T_{\rm eff}$ (Brown et al. 2018).
In Fig. 3 (cyan curve), we show the cumulative distance distribution for  the filtered sample. The median distance is 1.1~kpc and 95\% of the sample are within 2.1~kpc.

\begin{figure}[]
 \begin{center}
  \includegraphics[width=79mm,scale=1]{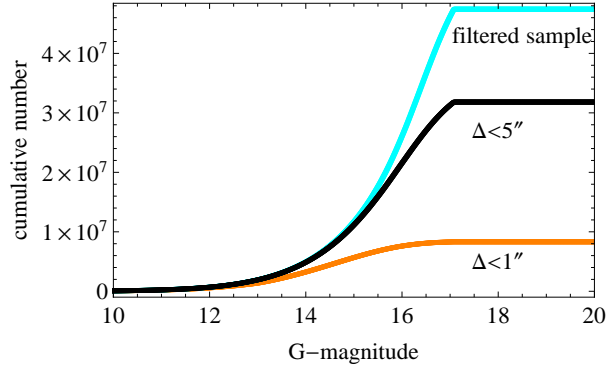}
   \caption{ Cumulative  G-magnitude  distributions of the filtered sample (cyan curve) and of the shooting targets with $\Delta <5$\rq\rq{} (black curve) and  $ <1$\rq\rq{} (orange curve).     }
 \end{center}
\end{figure}

\begin{figure}[]
 \begin{center}
  \includegraphics[width=79mm,scale=1]{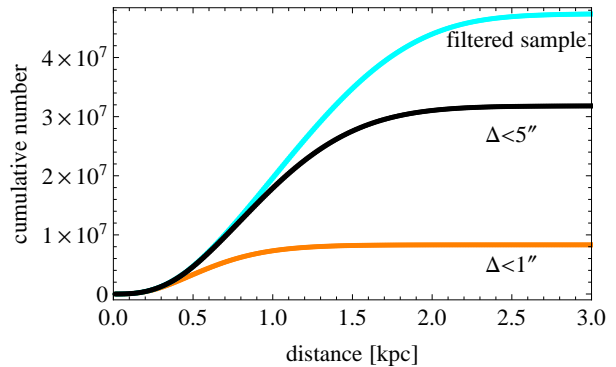}
   \caption{ Cumulative distance distributions of the filtered sample (cyan curve) and of the shooting targets with $\Delta <5$\rq\rq{} (black curve) and  $ <1$\rq\rq{} (orange curve).  The median distances are $ 0.57$~kpc and 0.92~kpc  for the orange and black curves.    }
 \end{center}
\end{figure}

Note that, for Gaia DR2, all stars are astrometrically analyzed as single stars, and some of the filtered sample   would be unfavorably  affected by the other members of multiple systems.  For multiple systems,  more elaborate analysis is planned   in the next Gaia data release,  and  we do not discuss the associated effects.

\subsection{selecting target stars}

\begin{figure}[t]
\vspace{7mm}
 \begin{center}
  \includegraphics[width=79mm,scale=1]{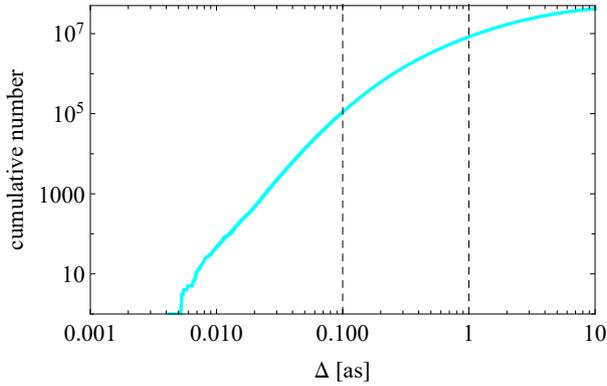}
   \caption{ Cumulative distributions of the uncertainty $\Delta $ expected for our filtered sample.   The two vertical dashed lines are at $\Delta= 0.1$\rq\rq{} and 1\rq\rq{}. }   
 \end{center}
\end{figure}

\begin{figure}[t]
 \begin{center}
  \includegraphics[width=79mm,scale=1]{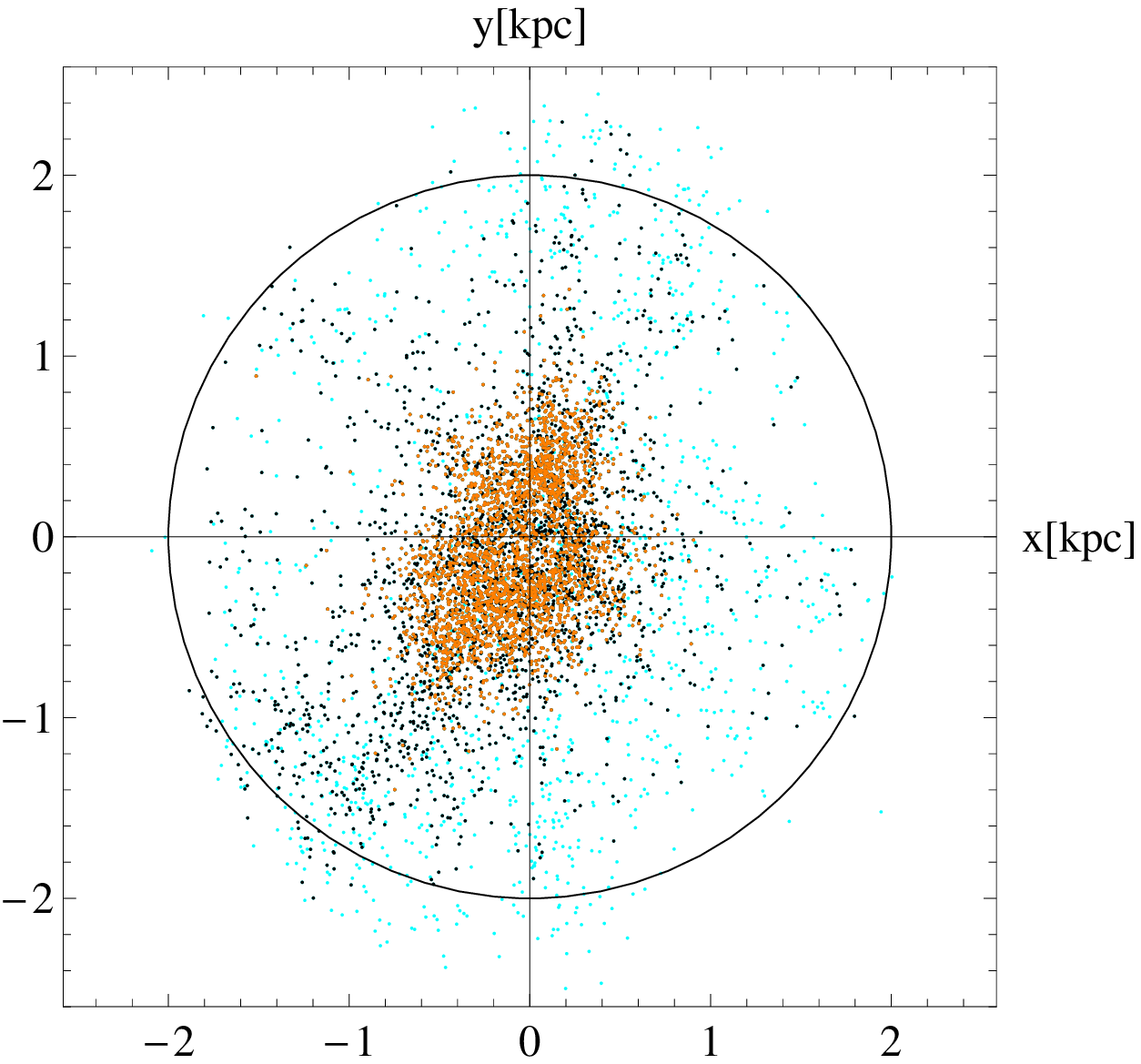}
\includegraphics[width=79mm,scale=1]{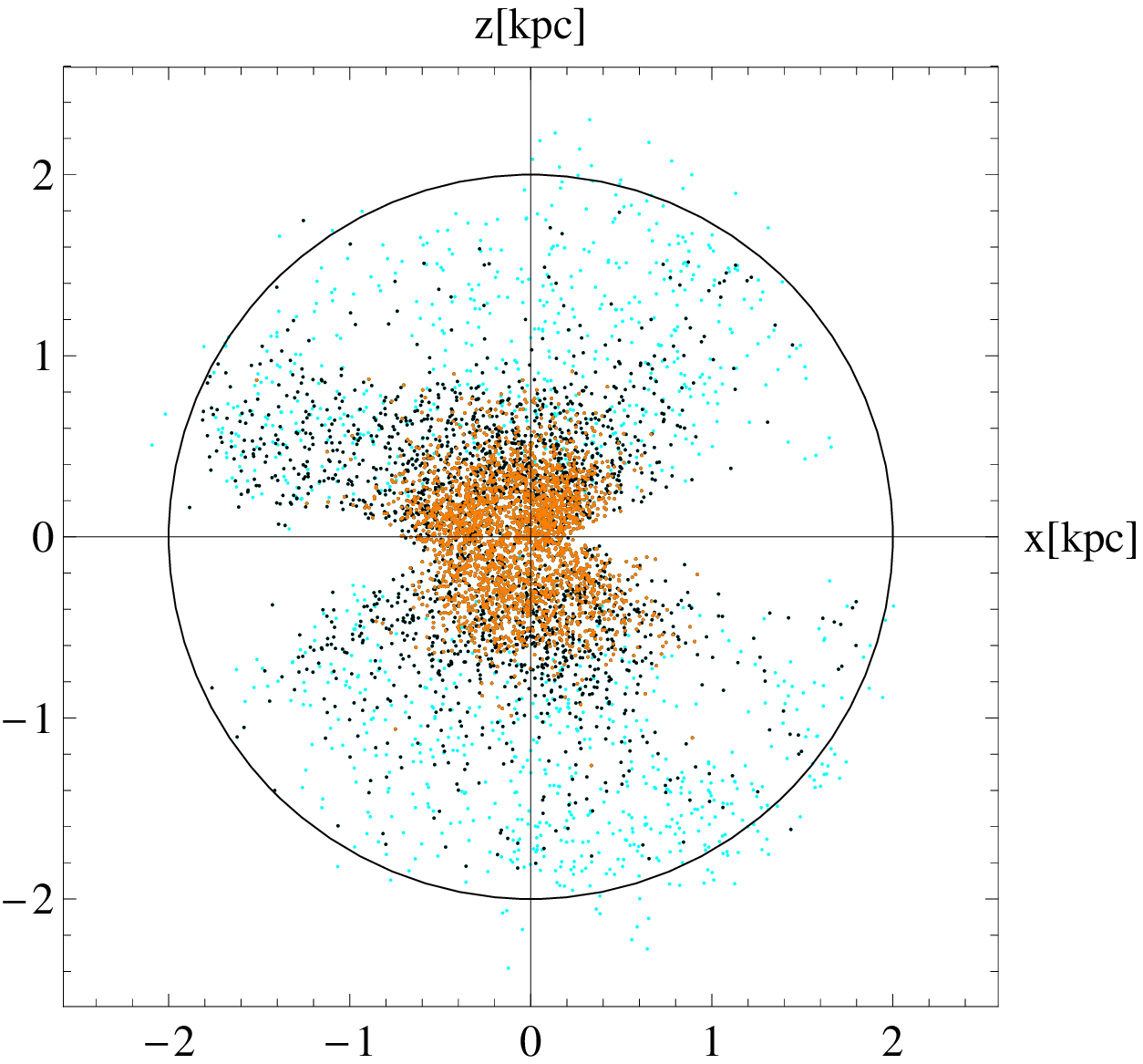}
   \caption{{\it  Upper panel}: Spatial distribution of 5928 solar-type stars with  $T_{\rm eff }\sim 5840$~K and $R\sim 1~R_\odot$.  All stars are projected to the Galactic ($xy$) plane. The orange dots represent 2738 target stars with $\Delta <1$\rq\rq{}  and the black ones show the additional 2109 target stars with 1\rq\rq{}$<\Delta <5$\rq\rq{}. The cyan dots are those with $\Delta>5$\rq\rq{}.     The Galactic center is toward the direction of $+x$-axis.  The radius of the black circle is 2~kpc.
{\it Lower panel}: Similarly projected to the $xz$-plane. The Galactic plane corresponds to the $x$-axis around which the number of stars are small due to the strong dust extinction.   }
 \end{center}
\end{figure}

Using the prescription based on the $2\times2$ covariance matrix (\ref{mat}) and the actual data provided in  Gaia DR2,  
we evaluate the angular uncertainty $\Delta$ for each of our filtered sample. In Fig. 4 we show the  cumulative distribution of $\Delta$. As expected from our order-of-magnitude estimation, the characteristic size is $O(1")$.  The best value is $\Delta=0.0042$\rq\rq{} for a K-star at the distance of 6.7~pc with $G=7.3$.  Given the incompleteness of the Gaia DR2 data at $G<12$, we should take the lower end of $\Delta $ just for a  reference as of now.

Now we
select our shooting targets by introducing the three fiducial  criterion values; $\Delta=0.1$\rq\rq{}, 1.0\rq\rq{} and 5.0\rq\rq{}, taking into account the specific numerical values quoted in introduction (for 0.1\rq\rq{} and 1.0\rq\rq{}).  The resulting numbers of the targets are summarized in Table 1.  We have $\Delta <5$\rq\rq{} for 67\% of the filtered sample, but the fractions decrease to 18\% ($\Delta <1$\rq\rq{}) and 0.2\% ($\Delta <0.1$\rq\rq{}) for more stringent requirements.

In Figs. 2 and 3, we show the distributions of G-magnitudes and distances for our targets with $\Delta <1$\rq\rq{} and 5\rq\rq{}. The median distances for the two criterion values are 0.57~kpc and 0.92~kpc, respectively.

To particularly  examine stars hosting confirmed planets, we   also utilize the cross match between Gaia DR2 and the NASA Exoplanet Archive. After applying the filters (i)-(iii) to totally 1678 cross-matched stars,  we obtain 1259 FGK stars as a subset of our filtered sample. We then evaluate their angular uncertainties $\Delta$ and obtain  26 targets with $\Delta <0.1$\rq\rq{}, 782 with $<1$\rq\rq{} and  1220 with $<5$\rq\rq{}.
If we limit our analysis only to host stars of confirmed habitable  planets, the subset size is reduced to 60 and we obtain
 1 target with $\Delta <0.1$\rq\rq{}, 48 with $<1$\rq\rq{} and  58 with $<5$\rq\rq{}.  Relative to the $4.7\times 10^7$ stars in our original sample, these two subsets have smaller uncertainties $\Delta$.  

\subsection{positions of the targets}

Here we discuss positions of our shooting  targets in the 
Galaxy. As a representative example, we specifically pick up the subset of the filtered sample whose 1-$\sigma$ error regions of $T_{\rm eff}$ and $R$ are simultaneously  within   $T_{\rm eff}\in [\rm 5790~K,5890~K]$ and $R\in[0.96,1.04]~R_\odot$, close to  the solar values. 
This subset contains 5928  solar-type stars, and we have 2738 targets for $\Delta<1$\rq\rq{}  and 4847  for $<5$\rq\rq{}. 
Note that the overall  data qualities  of this subset are better than our original filtered sample,  because of  the relatively strong requirements on $T_{\rm eff}$ and $R$.  Accordingly,  the target fractions of this subset  are higher than those in  Table 1.

For graphical demonstration,   we introduce a Cartesian coordinate $(x,y,z)$, using the distance $d$ and the Galactic angular coordinate $(l,b)$ as
\beq
(x,y,z)=d (\cos b \cos l,\cos b  \sin l,\sin b).
\eeq 
The Galactic center is at the direction of $+x$-axis, and the Galactic plane corresponds to the  $xy$-plane.
In the upper panel of Fig. 5, we show the projection of the subset sample onto the $xy$-plane. Most of the orange dots (targets with $\Delta <1$\rq\rq{}) have projected distances less than $\lsim1$~kpc, but the black dots (with 1\rq\rq{}$<\Delta <5$\rq\rq{})  are distributed over $\sim 2.0$~kpc. We expect that the observed anisotropy of the stars is mainly due to the Galactic extinction pattern and partially to the sampling pattern of Gaia.
{If we make a more detailed analysis,  we can identify a sparseness of the solar-type stars in the range $d\lsim 300$~pc. Given the absolute G-magnitude of the Sun $M_G=4.68$,  this is likely to be caused by the incomplete sampling of Gaia for bright stars at $G<12$.}

In the lower panel of Fig. 5, we show the projections of the stars onto the $xz$-plane. We can see a clear deficit of stars around the $x$-axis to which  the Galactic plane is projected. This reflects the strong extinction towards Galactic plane.  In future, some of the unaccessible volume might be explored by the proposed infrared missions such as  JASMINE (Gouda 2012) and GaiaNIR  (Hobbs et al. 2016). {Interestingly, along the $x$-axis,  the boundary of the orange points is not distinctively covered by the black points, unlike the $z$-axis direction.  This indicates  that the boundary is mainly determined by the limitation of the temperature estimation, not by the threshold value $\Delta=1$\rq\rq{}. }

When the Sun is observed inversely by ETI on a planet around a solar-type star  plotted in Fig. 4, they will record almost the same luminosity,  interstellar extinction and transverse velocity as we recorded for the star in Gaia DR2.  Therefore, if the ETI have astrometric mission equivalent to Gaia, they can realize a similar shooting accuracy $\Delta$ as we can expect for the star. Here we ignored   details such as the source density and orientation of their ecliptic plane, and also assumed that the extinction pattern does not change drastically below the arcmin scale. In this manner, Fig. 4 would be intriguing also from the view point of searching for intentional ETI signatures from solar-type stars, not just shooting them from the Earth.

\section{Discussions}

In this paper, we discussed the prospects of high precision pointing of our transmitters to Galactic habitable planets.  For a beam opening angle $\Theta$, we practically  want to estimate the future sky position of the host stars with accuracy better than  $\Delta<\Theta$.  This roughly  corresponds to measuring the transverse velocities $v_{\rm t}$ of the stars with a precision $\delta v_{\rm t}<0.75 {\rm~ km\,sec^{-1}} (\Theta/1")$ much smaller than the typical value $v_{\rm t}\sim 30 {\rm ~km\,sec^{-1}}$.  

 In the present  work, we regarded Gaia as an optimal instrument for our pointing problem.   Fully using the astrometric data provided in Gaia DR2, we evaluated the size of the angular uncertainties $\Delta$ individually for our filtered sample composed by $4.7\times 10^7$ FGK stars. 
As summarized in Table 1, we have the accuracy $\Delta<5$\rq\rq{} for 67\% of the filtered sample. The fraction decreases to 18\% for $\Delta <1$\rq\rq{}.

Until just a few years ago, Hipparcos catalog was the best available astrometric  data.  It includes $\sim 2\times 10^4$ stars whose distances were estimated within $\sim 10$\% errors.  As shown in  Eq. (8),  this  corresponds to the extrapolation error of at least $\sim 3" (v_{\rm t}/{\rm 30~km\, sec^{-1}})$.    With Gaia DR2, our target number is $O(10^7)$ for the accuracy goal  1.0\rq\rq{} (see Table 1),  and three orders of magnitude larger than  Hipparcos era.

 Gaia DR2 is based on the data collected in the first 22 months of observation. Gaia is smoothly operating now,  and the mission lifetime could be extended to $\sim 2024$, limited by  micro-propulsion
system fuel. \footnote{\url{https://www.cosmos.esa.int/gaia}} With 10 years data, ignoring instrumental degradation, the signal-to-noise ratio of the sources would increase by  a factor of 2.3, and the accuracies of the astrometric parameters would be improved  by at least the same  factor. The actual improvement is expected to be better than this simple scaling, considering the advantages of the long-term observation both on measuring  the proper motion   and on reducing  the noise correlation between the parallax and other parameters.    
If we conservatively use the improvement factor 2.3 for $\Delta$ in Fig. 4, the numbers of our shooting targets would be 2 and 7 times larger for $\Delta <1$\rq\rq{} and $<0.1$\rq\rq{} respectively, compared with Table 1.  In addition, as we commented earlier, information related to multiple stars would be refined.

In this paper, we studied interstellar communications mainly from the standpoint of a sender.  But a sender and a receiver are inextricably linked together, as demonstrated in Fig. 5.  In this sense,  our results would be suggestive also for SETI related activities.  Here, considering the rapid improvements  even of our technology  in the past few decades, it would be more productive to gain insights without making strong assumptions on the ETI\rq{}s technology level.

\ack[Acknowledgement]{
This work has made use of data from the European Space Agency (ESA)
mission Gaia, processed by
the {\it Gaia} Data Processing and Analysis Consortium (DPAC).
Funding
for the DPAC has been provided by national institutions, in particular
the institutions participating in the {\it Gaia} Multilateral Agreement.
This research was supported by the Munich Institute for Astro- and Particle Physics (MIAPP) which is funded by the Deutsche Forschungsgemeinschaft (DFG, German Research Foundation) under Germany\rq{}s Excellence Strategy EXC-2094-390783311.
We also used the gaia-kepler.fun crossmatch database created by Megan Bedell for the NASA Exoplanet Archive.
NS thanks H. Sugiura for his help on python codes.
 This work was supported by JSPS Kakenhi Grants-in Aid for Scientific Research (Nos. 17K14248,17H06358,18H04573,19K03870).}


\begin{thebibliography}{}


\bibitem[Andrae et al.(2018)]{2018A&A...616A...8A} Andrae, R., and 15 colleagues 2018.\ Gaia Data Release 2. First stellar parameters from Apsis.\ Astronomy and Astrophysics 616, A8.
\bibitem[Arnold(2013)]{2013IJAsB..12..212A} Arnold, L.\ 2013.\ Transmitting signals over interstellar distances: three approaches compared in the context of the Drake equation.\ International Journal of Astrobiology 12, 212.
\bibitem[Baum et al.(2011)]{2011AcAau..68.2114B} Baum, S.~D., Haqq-Misra, J.~D., Domagal-Goldman, S.~D.\ 2011.\ Would contact with extraterrestrials benefit or harm humanity? A scenario analysis.\ Acta Astronautica 68, 2114.

\bibitem[Benford et al.(2010)]{2010AsBio..10..475B} Benford, J., Benford, G., Benford, D.\ 2010.\ Messaging with Cost-Optimized Interstellar Beacons.\ Astrobiology 10, 475.
\bibitem[Clark and Cahoy(2018)]{2018ApJ...867...97C} Clark, J.~R., Cahoy, K.\ 2018.\ Optical Detection of Lasers with Near-term Technology at Interstellar Distances.\ The Astrophysical Journal 867, 97.

\bibitem[Cocconi and Morrison(1959)]{1959Natur.184..844C} Cocconi, G., Morrison, P.\ 1959.\ Searching for Interstellar Communications.\ Nature 184, 844.
\bibitem[De Simone et al.(2004)]{2004MNRAS.350..627D} De Simone, R., Wu, X., Tremaine, S.\ 2004.\ The stellar velocity distribution in the solar neighbourhood.\ Monthly Notices of the Royal Astronomical Society 350, 627.
\bibitem[Drake(1961)]{1961PhT....14...40D} Drake, F.~D.\ 1961.\ Project Ozma.\ Physics Today 14, 40.
\bibitem[Gaia Collaboration et al.(2016)]{2016A&A...595A...1G} Gaia Collaboration, and 625 colleagues 2016.\ The Gaia mission.\ Astronomy and Astrophysics 595, A1.




\bibitem[Gaia Collaboration et al.(2018)]{2018A&A...616A...1G} Gaia Collaboration, and 453 colleagues 2018.\ Gaia Data Release 2. Summary of the contents and survey properties.\ Astronomy and Astrophysics 616, A1.
\bibitem[Gajjar et al.(2019)]{2019BAAS...51g.223G} Gajjar, V., and 36 colleagues 2019.\ The Breakthrough Listen Search for Extraterrestrial Intelligence.\ Bulletin of the American Astronomical Society 51, 223.
\bibitem[Gouda(2012)]{2012ASPC..458..417G} Gouda, N.\ 2012.\ Infrared Space Astrometry Missions JASMINE Missions.\ Galactic Archaeology: Near-field Cosmology and the Formation of the Milky Way 417.
\bibitem[Guillochon and Loeb(2015)]{2015ApJ...811L..20G} Guillochon, J., Loeb, A.\ 2015.\ SETI via Leakage from Light Sails in Exoplanetary Systems.\ The Astrophysical Journal 811, L20.
\bibitem[Hippke(2019)]{2019arXiv191202616H} Hippke, M.\ 2019.\ Interstellar communication network. I. Overview and assumptions.\ arXiv e-prints arXiv:1912.02616.
\bibitem[Hobbs et al.(2016)]{2016arXiv160907325H} Hobbs, D., and 24 colleagues 2016.\ GaiaNIR: Combining optical and Near-Infra-Red (NIR) capabilities with Time-Delay-Integration (TDI) sensors for a future Gaia-like mission.\ arXiv e-prints arXiv:1609.07325.
\bibitem[Horowitz and Sagan(1993)]{1993ApJ...415..218H} Horowitz, P., Sagan, C.\ 1993.\ Five Years of Project META: an All-Sky Narrow-Band Radio Search for Extraterrestrial Signals.\ The Astrophysical Journal 415, 218.
\bibitem[Perryman et al.(2014)]{2014ApJ...797...14P} Perryman, M., Hartman, J., Bakos, G. {\'A}., Lindegren, L.\ 2014.\ Astrometric Exoplanet Detection with Gaia.\ The Astrophysical Journal 797, 14.
\bibitem[Petigura et al.(2013)]{2013PNAS..11019273P} Petigura, E.~A., Howard, A.~W., Marcy, G.~W.\ 2013.\ Prevalence of Earth-size planets orbiting Sun-like stars.\ Proceedings of the National Academy of Science 110, 19273.
\bibitem[Schelling(1960)]{}Schelling, T. C. 1960, The strategy of conflict (Cambridge, Massachusetts: Harvard University Press)
\bibitem[Seto(2019)]{2019ApJ...875L..10S} Seto, N.\ 2019.\ Possibility of a Coordinated Signaling Scheme in the Galaxy and SETI Experiments.\ The Astrophysical Journal 875, L10.
\bibitem[Siemion et al.(2013)]{2013ApJ...767...94S} Siemion, A.~P.~V., and 10 colleagues 2013.\ A 1.1-1.9 GHz SETI Survey of the Kepler Field. I. A Search for Narrow-band Emission from Select Targets.\ The Astrophysical Journal 767, 94.
\bibitem[Tarter et al.(2010)]{2010SPIE.7819E..02T} Tarter, J.~C., and 13 colleagues 2010.\ SETI turns 50: five decades of progress in the search for extraterrestrial intelligence.\ Instruments, Methods, and Missions for Astrobiology XIII 781902.
\bibitem[Tarter(2001)]{2001ARA&A..39..511T} Tarter, J.\ 2001.\ The Search for Extraterrestrial Intelligence (SETI).\ Annual Review of Astronomy and Astrophysics 39, 511.
\bibitem[Titov and Lambert(2013)]{2013A&A...559A..95T} Titov, O., Lambert, S.\ 2013.\ Improved VLBI measurement of the solar system acceleration.\ Astronomy and Astrophysics 559, A95.
\bibitem[Vakoch(2016)]{2016NatPh..12..890V} Vakoch, D.~A.\ 2016.\ In defence of METI.\ Nature Physics 12, 890.
\bibitem[Wright et al.(2018)]{2018AJ....156..260W} Wright, J.~T., Kanodia, S., Lubar, E.\ 2018.\ How Much SETI Has Been Done? Finding Needles in the n-dimensional Cosmic Haystack.\ The Astronomical Journal 156, 260.
\bibitem[Wright(2018)]{2018haex.bookE.186W} Wright, J.~T.\ 2018.\ Exoplanets and SETI.\ Handbook of Exoplanets 186.
\bibitem[Zaitsev(2016)]{2010p} Zaitsev, P. 2010, in Searching for extraterrestrial intelligence, ed. H. P. Shuch (Chichester, UK: Springer), 399




\end{thebibliography}
\end{document}